\documentstyle[sprocl,psfig]{article}

\def\Journal#1#2#3#4{{#1} {\bf #2}, #3 (#4)}

\def\be{\begin{equation}}
\def\ee{\end{equation}}
\def\bea{\begin{eqnarray}}
\def\eea{\end{eqnarray}}

\def\Fig#1{Fig.\ \ref{fg:#1}}	

\def\farcm{\hbox{$.\mkern-4mu^\prime$}}
\def\farcs{\hbox{$.\!\!^{\prime\prime}$}}
\def\gtrsim{\mathrel{\hbox{\rlap{\hbox{\lower4pt\hbox{$\sim$}}}\hbox{$>$}}}}
\def\hkpc{\mbox{$h^{-1}\ {\rm kpc}$}}

\def\kpc{\mbox{$\ {\rm kpc}$}}
\def\lesssim{\mathrel{\hbox{\rlap{\hbox{\lower4pt\hbox{$\sim$}}}\hbox{$<$}}}}
\def\lya{\mbox{${\rm Ly}\alpha$}}
\def\mpc{\mbox{$\ {\rm Mpc}$}}

\let\la=\lesssim			
\let\ga=\gtrsim

\begin{document}

\title{Galaxies at High Redshifts}

\author{A. Yahil, K. M. Lanzetta}

\address{Department of Physics and Astronomy, State University of New York,\\
Stony Brook, NY 11794-3800, USA}

\author{A. Fern\'andez-Soto}

\address{Department of Astrophysics and Optics, School of Physics, University
of New South Wales, Kensington-Sidney, NSW 2052, Australia}


\maketitle\abstracts{Several conclusions have been reached over the last few
years concerning high-redshift galaxies: (1)~The excess of faint blue galaxies
is due to dwarf galaxies.  (2)~Star formation peaks at redshifts $z\approx
1-2$.  (3)~It appears to occur piecemeal in any given galaxy and there is no
evidence for starbursting throughout a large $\sim 10\kpc$ galaxy.  (4)~There
is significant and sharp diminution in the number of $L_\star$ spiral galaxies
at redshifts $1<z<2$ and elliptical galaxies at redshifts $2.5<z<4$. (5)~It is
increasingly more difficult to ``hide'' large high-redshift galaxies in
universes with larger volumes per unit redshift, i.e., open or $\lambda$
models, which have lower deceleration parameters.}
  
\section{Introduction}

This paper reviews high redshift galaxies as we understand them.  It is not
meant to be comprehensive.  Instead, we ask the questions that seem most
important to us.  Why do we observe high redshift galaxies, \S\ref{sec:why}?
How do we observe them, \S\ref{sec:how}?  What have we learned from the
observations, \S\ref{sec:learned}?  And how should we continue our research,
\S\ref{sec:where}?

\section{Why Observe High Redshift Galaxies?\label{sec:why}}

In the standard paradigm of galaxy formation, galaxies are imagined to have
formed in essentially their present shapes at some definite epoch in the past,
at redshifts $z\sim 5$.  Their stellar content is then supposed to evolve
passively with time, governed only by the rate of star formation and stellar
evolution.  Elliptical galaxies are supposed to have a short initial
star-formation phase $\sim 1$ Gyr, while spirals and irregulars have star
formation lasting 10 Gyr or longer.

The original motivation for passive-evolution was the observation of increased
counts of blue galaxies at magnitudes $B\ga 24$, which were thought to be
high-redshift elliptical galaxies in their original burst of star
formation.\cite{tinsley80}

Today we are able directly to observe high-redshift galaxies and hence see the
formation and evolution of both elliptical and spiral galaxies in action;
evolution can be tested directly.  Although less discussed, observations have
now been carried out sufficiently deeply to have a chance to observe the
initial burst of star formation even in individual globular clusters.  More
generally, objects whose luminosities are well below $L_\star$ are now
routinely observed at high redshift.

With systematic deep surveys it has also become possible to estimate the total
rest-frame UV radiation per unit volume emitted at high redshift and to deduce
the global rate of star formation.\cite{madau96}  These observations can then
be compared with simulations of galaxy formation.\cite{baugh98}

\begin{figure}[h]
\psfig{figure=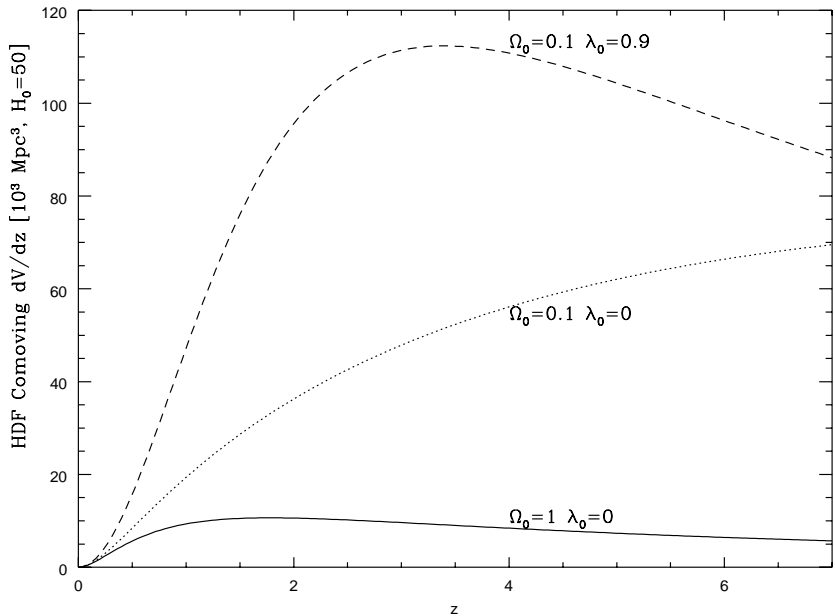,width=\textwidth,angle=-90}
\caption{\label{fg:dvol} Differential comoving volume versus redshift in the
Hubble Deep Field for three cosmological models.  Note the order-of-magnitude
difference between the models.}
\end{figure}

Finally, the counts of galaxies as a function of redshift, luminosity, and, as
we argue below, size, can be used to constrain the deceleration parameter,
$q_0$.  This is a difficult task, due to uncertainties in galaxy formation and
evolution and in the dust content of young galaxies.  At redshifts $z\ga 1$,
though, the difficulty is mitigated by the order-of-magnitude sensitivity of
the differential comoving volume, $dV/dz$, to the deceleration parameter,
\Fig{dvol}.

\section{How to Observe High Redshift Galaxies?\label{sec:how}}

The attempt to detect primeval galaxies by means of strong emission lines,
particularly \lya, has failed.  Blind surveys for emission-line objects have
covered more than $10^6 \mpc^3$ with no detections.\cite{thompson95}  We now
understand that even a minuscule amount of dust can strongly attenuate the
\lya\ photons that are resonantly trapped in the emitting region.

High redshift galaxies are occasionally discovered serendipitously, for example
when they are lensed by a foreground cluster.\cite{frye98} More recently, a
possible rich cluster of galaxies was identified at $z=2.56$ in a field in
which there are two confirmed quasars of that redshift separated by 1\farcm5,
and a nearby microwave background decrement that is plausibly explained as due
to the Sunyaev-Zel'dovich effect.\cite{campos98} The cluster candidates, which
await spectroscopic confirmation, were identified by means of excess emission
at the redshifted \lya\ wavelength detected by medium-band imaging.

While such discoveries are of interest --- even a single confirmed rich cluster
at high redshift could significantly constrain some cosmological models ---
real understanding comes only with systematic surveys based on a successful
detection scheme.  Such a mechanism has become available only in the last few
years, with the realization that absorption at the Lyman limit and in the
\lya-forest region lead to broad-band color features which provide reliable
photometric redshift estimates.\cite{steidel92}

The need for photometric redshifts became acute as the limiting imaging
magnitude was pushed to fainter limits than are accessible to spectroscopy,
particularly the faint galaxies of the Hubble Deep Field (HDF).  The methods
for photometry and redshift estimates vary somewhat between different groups,
but there is now agreement on the resultant redshifts, and they are in good
agreement with spectroscopic redshifts where available.\cite{hogg98} In fact,
the spectroscopy is not easy either, and under careful scrutiny more
spectroscopic redshifts were found to be in error than photometric
ones.\cite{lanzetta98a}

\section{What Have We Learned?\label{sec:learned}}

\begin{figure}[h]
\psfig{figure=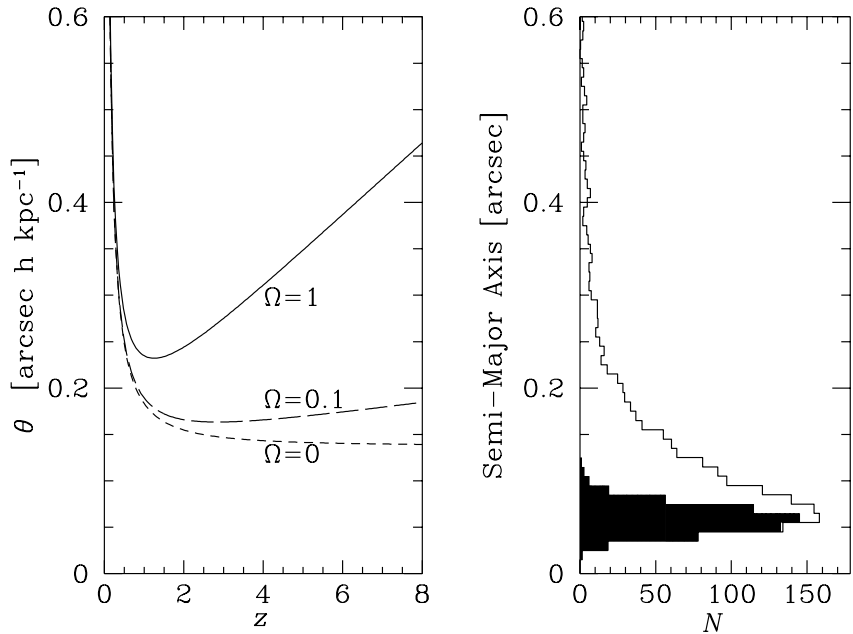,width=\textwidth,angle=-90}
\caption{\label{fg:ahist} The expected relation between angular diameter and
redshift for an object with proper size $1\hkpc$, compared with the actual
distribution of diameters of galaxies in the Hubble Deep Field.  The shaded
part of the histogram are the counts of galaxies with magnitudes $AB(8140) >
28$ where incompleteness sets it.}
\end{figure}

The first and foremost fact to emerge from the imaging of the HDF was the small
angular size of the bulk of the galaxies.  The right panel of \Fig{ahist} shows
the angular distribution of all the HDF objects, most of which are $\sim
0\farcs 1$.  As the left panel of the figure shows, this corresponds to a size
$\la 1 \kpc$, and an inspection of the HDF images shows that most of them are
blue.  Hence, the excess faint blue galaxies are dwarfs, not large,
high-redshift, elliptical galaxies.

\begin{figure}[h]
\psfig{figure=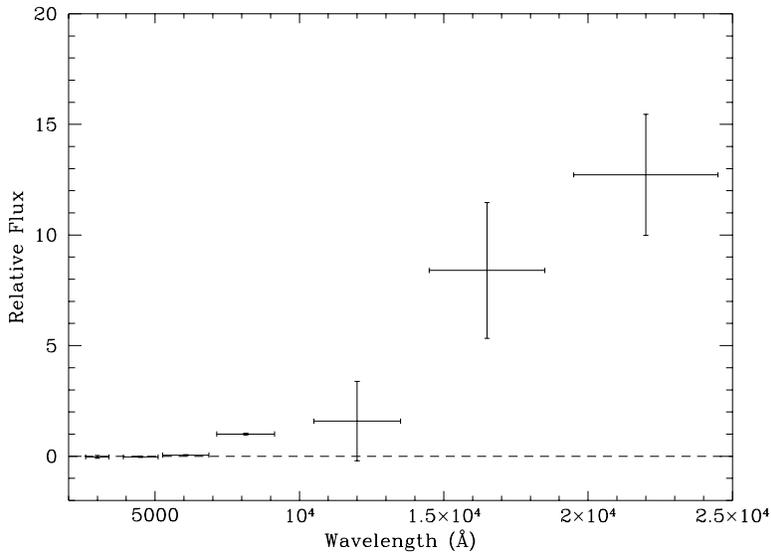,width=\textwidth,angle=-90}
\caption{\label{fg:z6} Spectral energy distribution of a galaxy with
photometric redshift $z>5$.  Note the small sizes of the error bars of the HST
observations, which are almost imperceptible on the scale plotted but severely
constrain the spectral energy distributions that can be fitted to the data.
The redshift estimate is based on the sharp drop in flux between 6000\AA\ and
8000\AA, which is attributed to Lyman-limit and \lya-forest absorption.
Absorption by dust is ruled out because it would imply higher infrared flux
than is observed.}
\end{figure}

Among the HDF galaxies are a handful of galaxies at photometric redshifts $z
\ga 5$,\cite{lanzetta96,lanzetta98b} the brightest of which are not much
fainter than those observed spectroscopically.  \Fig{z6} shows an example of
the spectral energy distribution of such a galaxy with no detection shortward
of the F814W band, clear detection in the F814W and K bands and a probable one
in the H band.  Notice the small sizes of the error bars of the HST
observations, which are almost imperceptible on the scale plotted.  They
severely constrain the spectral energy distributions that can be fitted to the
data.  The lower limit on the F814W/F606W flux ratio can also not be due to
dust (in extraordinarily large amounts) because the spectrum would rise much
more steeply into the IR, exceeding the observations.  Finally,
multiple-wavelength detection rules out the possibility of an emission-line
galaxy, and the angular extension precludes a star.  The only plausible
explanation of the sharp drop in flux between the F814W and F606W bands is
Lyman-limit and \lya-forest absorption, placing the galaxy at redshift $z \ga
5$.

\begin{figure}[h]
\psfig{figure=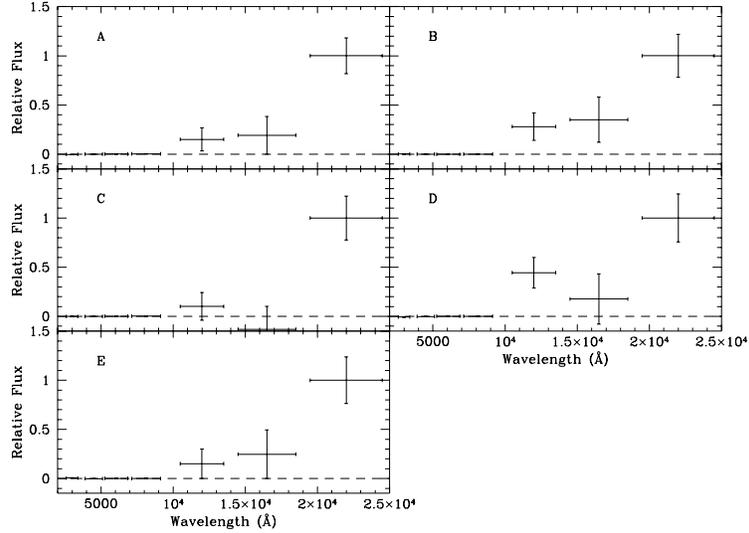,width=\textwidth,angle=-90}
\caption{\label{fg:z20_spectrum} Spectral energy distributions of five objects
detected in ground-based $K$-band imaging but not in the HST F814W band.
Interpreting the drop in flux in the HST bands to be due to Lyman-limit and
\lya\ absorption, these objects have redshifts in the range $z=7-17$.}
\end{figure}

The search for high-redshift galaxies can be extended further by looking for
galaxies which show IR emission but no detection in any of the HST bands,
including the F814W band.  \Fig{z20_spectrum} shows the spectral energy
distributions of five objects detected in the K band but not in the F814W band.
If the drop in flux in the F814W band is due to Lyman-limit and \lya\
absorption, then the redshifts of these objects are in the range $z=7-17$; a
more accurate determination awaits better photometry, particularly in the J and
H bands.

\begin{figure}[h]
\psfig{figure=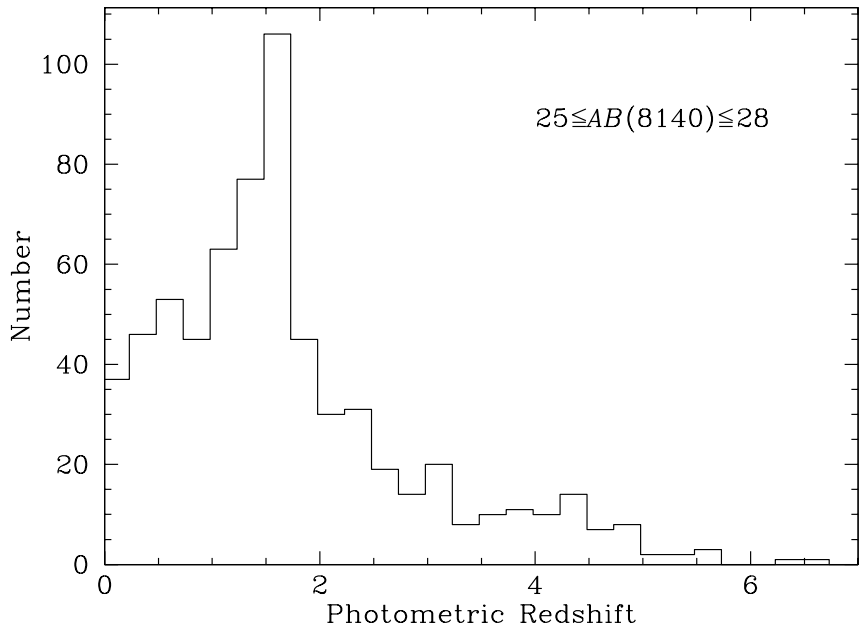,width=\textwidth,angle=-90}
\caption{\label{fg:nz} Distribution of photometric redshifts of all objects
with magnitudes $25\le AB(8140)\le 28$ in the Hubble Deep Field.  Note the
decline in numbers for redshifts $z\ga 2.5$ and contrast it with the expected
increase for constant comoving density, \Fig{dvol}.}
\end{figure}

Although it is very exciting to detect galaxies at higher redshifts than
previously known, the most important ingredient to our understanding of galaxy
formation and evolution comes not from those exceptional tails of the
distribution but from the overall redshift distribution.  \Fig{nz} shows the
photometric redshift distribution, whose most remarkable feature is the drop in
counts at redshifts $z \ga 2.5$.  A morphological breakdown of the redshift
distribution shows, in fact, that spiral galaxies disappear quickly beyond
$z=1$, and elliptical galaxies beyond $z=2.5$.\cite{driver98} All the objects
at higher redshifts are small, with sizes $\sim 1\kpc$.

Recall that high-redshift galaxies are observed at rest-UV wavelengths, and
that the UV emission in nearby galaxies is also confined to starbursting
regions of comparable size.\cite{meurer95} It is therefore tempting to explain
the small sizes of high-redshift galaxies as due to UV-producing starbursts in
small regions of galaxies and not over an entire 10\kpc\ range.

However, the drop in the number counts of 10\kpc\ spirals at redshift $z\approx
1$, and that of the ellipticals at redshift $z\approx 2.5$, is quite sharp.
The HDF is deep enough so galaxies with luminosities well below $L_\star$ can
be detected.  It can also be shown that the cosmological diminution of surface
brightness, together with the K-correction, are not the limiting factors for
redshifts $z\la2$ (spirals) and $z\la4$ (ellipticals).  Present day galaxies,
evolved according to standard models,\cite{bruzual93} would be observed to have
larger sizes than the objects in the HDF.  However, uncertainties in the
evolutionary models, plus the unknown dust content of primeval galaxies, make
it difficult to reach a firm conclusion.

It is clear that the redshift cutoff cannot be explained by having star
formation take place entirely at lower redshifts, since this would violate
observational limits from number counts and redshift surveys.  If the galaxies
were indeed small at high-redshift, they have had to grow by mergers
(ellipticals), or star formation in their disks has had to proceed from the
inside out (spirals).

Whatever the mechanism by which galaxies ``hide'' at high redshift, it is
increasingly more difficult to conceal them in large-volume universes, such as
open or $\lambda$ universes with a low deceleration, \Fig{dvol}.  For those
cosmological models the total number of observed galaxies is less than
predicted, irrespective of size.  Star formation would then have to be episodal
to account for the observed number of objects, and more of the star formation
would have to take place behind a shroud of dust.  Any realistic model of
galaxy formation will have to face this issue, particularly if other
observational evidence continues to point toward open or $\lambda$ models, as
it now increasingly does.

Further comments are given in the concluding remarks to this conference.

\section{Where Do We Go From Here?\label{sec:where}}

Based on the experience of the last few years we expect most progress to come
from observations.  Deep near-infrared imaging is very useful to improve the
photometric redshifts of all fainter galaxies, and is essential for galaxies at
redshifts $z \ga 5$.  Spectroscopy of Balmer lines can determine the dust
content of the galaxies, telling us how much star formation is hidden from UV
view.  It will also confirm spectroscopic redshifts in the range $1 < z < 2$,
which have not been adequately observed from the ground.  Deep HST imaging in
intermediate bands, especially at 7000\AA, can improve the photometric
redshifts and increase their reliability at fainter magnitudes.  Finally,
better image reconstruction should improve both resolution and photometry,
enabling more accurate flux measurement and the detection of fainter objects.

On the theoretical side, modeling of galaxy formation is only beginning
seriously to tackle gas dynamics and star formation.  Expect important
breakthroughs here.

In any event, we can no longer content ourselves with the traditional number
counts, $N(m)$, $N(z)$, or even $N(m,z)$.  The sizes of galaxies at different
wavelengths, or better yet their profiles, are observable and provide crucial
constraints both to galaxy formation and to the underlying cosmological model.
We need to obtain these data and to confront our models with them.

\section*{Acknowledgments}

This work was supported in part by NASA grant AR-07551.01-96A.


\section*{References}
\begin{thebibliography}{99}

\bibitem{tinsley80} B. M. Tinsley \Journal{ApJ}{241}{41}{1980}.

\bibitem{madau96} P. Madau, H. C. Ferguson, M. E. Dickinson, M. Giavalisco,
C. C. Steidel, and A. Fruchter, \Journal{MNRAS}{283}{1388}{1996}.

\bibitem{baugh98} For example, C. M. Baugh, S. Cole, C. S. Frenk, and
C. G. Lacey, ApJ, in press (1998); preprint astro-ph/9703111.

\bibitem{thompson95} D. Thompson and S. G. Djorgovski,
\Journal{AJ}{110}{982}{1995}.

\bibitem{frye98} B. Frye and T. Broadhurst, ApJ, submitted, and references
therein; preprint astro-ph/9712111.

\bibitem{campos98} A. Campos, A. Yahil, R. A. Windhorst, E. A. Richards, and
S. Pascarelle, ApJ, submitted (1998).

\bibitem{steidel92} C. C. Steidel and D. Hamilton,
\Journal{AJ}{104}{941}{1992}.

\bibitem{hogg98} D. W. Hogg, et al., AJ, in press (1998); preprint
astro-ph/9801133.

\bibitem{lanzetta98a} K. M. Lanzetta, A. Fern\'andez-Soto, and A. Yahil, in
{\em The Hubble Deep Field\/}, eds. M. Livio, S. M. Fall and P. Madau, in
press; preprint astro-ph/9709166.

\bibitem{lanzetta96} K. M. Lanzetta, A. Yahil, and A. Fern\'andez-Soto,
\Journal{Nature}{381}{759}{1996}.

\bibitem{lanzetta98b} K. M. Lanzetta, A. Fern\'andez-Soto, and A. Yahil, AJ,
submitted (1998)

\bibitem{driver98} S. P. Driver, A. Fern\'andez-Soto, W. J. Couch,
S. C. Odewahn, R. A. Windhorst, S. Phillipps, K. Lanzetta, and A. Yahil, ApJ,
in press; preprint astro-ph/9802092.

\bibitem{meurer95} G. R. Meurer, T. M. Heckman, C. Leitherer, A. Kinney,
C. Robert, and D. R. Garnett, \Journal{AJ}{110}{2665}{1995}.

\bibitem{bruzual93} G. Bruzual and S. Charlot, \Journal{ApJ}{405}{538}{1993};
and later releases of computer codes.

\end{thebibliography}
\end{document}